\documentclass[12pt,a4paper,DIV12]{scrartcl}
\usepackage[utf8x]{inputenc}
\usepackage[T1]{fontenc}
\usepackage{lmodern}
\usepackage[british]{babel}
\usepackage{amsmath}
\usepackage{amssymb}
\usepackage{amsfonts}
\usepackage{graphicx}
\usepackage{hyperref}
\usepackage{xcolor}
\usepackage{subfigure}
\usepackage{authblk}
\usepackage{braket}

\newcommand{\arxiv}[2]{[arXiv:\,\href{http://arxiv.org/abs/#1}
{\texttt{#1}}[\texttt{#2}]]}
\newcommand{\arxivold}[1]{[arXiv:\,\href{http://arxiv.org/abs/#1}
{\texttt{#1}}\,]}
\newcommand{\I}{\ensuremath{\mathrm{i}}}
\newcommand{\E}{\ensuremath{\mathrm{e}}}
\newcommand{\tr}{\ensuremath{\mathrm{tr}}}
\newcommand{\mg}{m_{\tilde{g}}}
\newcommand{\mgg}{m_{g\tilde{g}}}
\newcommand{\api}{\ensuremath{\text{a-}\pi}}
\newcommand{\aetap}{\ensuremath{\text{a-}\eta'}}

\title{%
  {\vspace{-20mm}\normalsize
   \hfill\parbox[b][30mm][t]{35mm}{\textmd{MS-TP-23-05}}}\\[-18mm]
Baryonic states in $\mathbf{\mathcal{N}=1}$ supersymmetric SU(2) Yang-Mills
theory on the lattice
}
\author[1,2]{Sajid Ali\thanks{sajid.ali@physik.uni-bielefeld.de}}
\author[3,4]{Georg Bergner\thanks{georg.bergner@uni-jena.de}}
\author[3]{Camilo L\'opez\thanks{camilo.lopez@uni-jena.de}}
\author[5]{Istvan Montvay\thanks{montvay@mail.desy.de}}
\author[4]{Gernot M\"unster\thanks{munsteg@uni-muenster.de}}
\author[6]{Stefano Piemonte\thanks{stefano.piemonte@ur.de}}
\affil[1]{Universit\"at Bielefeld, Fakult\"at f\"ur Physik,
Universit\"atsstr.~25, D-33615 Bielefeld, Germany}
\affil[2]{Government College University Lahore, Department of Physics,
Lahore 54000, Pakistan}
\affil[3]{University of Jena, Institute for Theoretical Physics, 
Max-Wien-Platz 1, D-07743 Jena, Germany}
\affil[4]{University of M\"unster, Institute for Theoretical Physics, 
Wilhelm-Klemm-Str.~9, D-48149 M\"unster, Germany}
\affil[5]{Deutsches Elektronen-Synchrotron DESY, Notkestr.~85, D-22607
Hamburg, Germany}
\affil[6]{University of Regensburg, Institute for Theoretical Physics, 
Universit\"atsstr.~31, D-93040 Regensburg, Germany}
\date{\today}
\begin{document}
\maketitle

\newpage

\begin{abstract}
\noindent
\textbf{\textsf{Abstract:}} 
We extend our analysis of bound states in $\mathcal{N}=1$ supersymmetric Yang-Mills 
theory by the consideration of baryonic operators, which are composed of three gluino 
fields. The corresponding states are similar to the baryons in QCD, but due to the 
difference between gluino and quark fields, their properties and the fermion line 
contractions involved in their correlation functions are different from QCD. In this work, 
we first explain the derivation of these operators and the contractions needed in numerical 
calculations of their correlators. In contrast to QCD the correlators contain a spectacle 
piece, which requires methods for all-to-all propagators. We provide a first estimate 
of the two-point function and the mass of the lightest baryonic state in $\mathcal{N}=1$ 
supersymmetric Yang-Mills theory.
\end{abstract}

\section{Introduction}

Supersymmetry (SUSY) provides field theoretic models, which are interesting in view
of various aspects of elementary particle theory. Supersymmetric extensions of
the Standard Model are able to resolve the hierarchy problem
\cite{Lykken:2010mc}, and they include dark matter candidates \cite{Jungman:1995df}.
Supersymmetry enforces structural properties on models that can be investigated
by perturbative or nonperturbative methods. This article addresses the
$\mathcal{N}=1$ supersymmetric Yang-Mills (SYM) theory, which represents
the supersymmetric extension of the gluonic sector of the Standard Model  
\cite{Amati:1988}. Gluons are described as usual by non-abelian gauge
fields $A^a_\mu(x)$ for gauge group SU($N_c$), where $a=1, \ldots, N^2_c-1$.
In addition to the gluons, SYM theory contains gluinos as their
superpartners. Gluinos are Majorana fermions transforming under the adjoint
representation of the gauge group. They are described by gluino fields 
$\lambda^a(x)$. In Minkowski space, the on-shell Lagrangian for 
$\mathcal{N}=1$ SYM theory, describing strongly interacting gluons and
gluinos, is given by
\begin{equation}
\mathcal{L}_{\text{SYM}} = -\frac{1}{4} F^a_{\mu\nu} F^{a,\mu\nu} 
+ \frac{\I}{2} \bar{\lambda}^a \gamma^\mu \left( \mathcal{D}_\mu \lambda \right)^a 
- \frac{\mg}{2} \bar{\lambda}^a \lambda^a .
\end{equation}
Here $F^a_{\mu\nu}$ is the non-abelian field strength tensor, and $\mathcal{D}_\mu$
is the covariant derivative in the adjoint representation of the gauge group.
The Lagrangian also includes a gluino mass term with mass
$\mg$. For $\mg \neq 0$ this term breaks SUSY softly, 
which means that it does not affect the renormalisation properties of the
theory and that the spectrum of the theory depends on
the gluino mass in a continuous way. 

In our previous investigations of SYM theory, we have concentrated on the
low-lying mass spectrum of the theory with gauge group SU(2) and SU(3),
which we have calculated nonperturbatively from first principles using Monte Carlo
techniques \cite{Bergner:2015adz,Ali:2017iof,Ali:2018dnd,Ali:2019gzj,Ali:2019agk}.
In addition, we have studied the SUSY Ward identities \cite{Ali:2018fbq,Ali:2020mvj}.
The particle spectrum of SYM theory is expected to consist of color neutral
bound states of gluons and gluinos, which should form mass degenerate
supermultiplets, if SUSY is not broken \cite{Veneziano:1982ah,Farrar:1997fn}.
In our numerical calculations, extrapolated to the continuum limit, we indeed
obtain mass degenerate supermultiplets \cite{Ali:2019agk}.

The predictions of \cite{Veneziano:1982ah,Farrar:1997fn} for the low-lying
supermultiplets are based on effective Lagrangeans, which describe
bound states of two gluinos, bound states of a gluon and a gluino, and glueballs.
Our previous numerical calculations have been focused on these types of
particles. 
Due to the fact that gluinos are in the adjoint representation of the gauge group, 
it is, however, also possible for any number $N_c$ of colors to form color neutral 
bound states of three gluinos. As they are analogous to the baryons of QCD, we call
these bound states generally ``baryons'', even for gauge group SU(2), although
bound states of $N_c$ fermions would commonly be called baryons.

Baryonic states in SYM theory have so far not being considered in the literature.
It is the aim of this article to describe the theoretical framework for a
numerical study of baryons in SYM theory, and to present the results of an
explorative calculation.

Related baryonic states have been investigated in SU(2) Yang-Mills theory coupled 
to one Dirac fermion in the adjoint representation \cite{Bi:2019gle} with a different
motivation from our study. In this case, there are conjectures about baryonic fields as 
dominant low energy degrees of freedom \cite{Anber:2018iof}.

For the Monte-Carlo simulations on a Euclidean four-dimensional
hypercubic lattice we use the action proposed by Curci and Veneziano
\cite{Curci:1986sm}. The gauge part $S_g$ of the complete action
$S = S_g + S_f$ is the usual plaquette action
\begin{equation}
S_g = -\frac{\beta}{N_{c}} \sum_{p} 
\mathrm{Re} \left[ \tr \left( U_{p} \right) \right],
\end{equation}
with the inverse gauge coupling given by $\beta = 2N_c/g^2$. In the
fermionic part $S_f$ the gluinos are implemented as Wilson fermions:
\begin{align}
S_f &=
\frac{1}{2} \sum_{x} \left\{ \bar{\lambda}^{a}_{x} \lambda_{x}^{a} 
- \kappa \sum_{\mu = 1}^{4} 
\left[ \bar{\lambda}^{a}_{x + \hat{\mu}} V_{ab, x \mu} (1 + \gamma_{\mu}) 
\lambda^{b}_{x}
+ \bar{\lambda}^{a}_{x} V^{T}_{ab, x \mu} (1 - \gamma_{\mu})
\lambda^{b}_{x + \hat{\mu}} \right] \right\} \\
&\equiv \frac{1}{2} \sum_{x,y} \bar{\lambda}^{a}_{x} D_w^{ab}(x,y)
\lambda_{y}^{b} ,
\end{align}
where $D_w$ is the Wilson-Dirac matrix.
The link variables in the adjoint representation are given by
$V_{ab, x \mu} = 2\,\tr\,(U_{x\mu}^\dagger T_a U_{x\mu} T_b)$, where $T_a$
are the generators of the gauge group.
The hopping parameter $\kappa$ is related to the bare gluino mass $\mg$
by $\kappa = 1/(2 \mg + 8)$. In order to approach the limit of vanishing
gluino mass, the hopping parameter has to be tuned properly.
In our numerical investigations the fermionic part is additionally
$O(a)$ improved by adding the clover term
$-(c_{sw}/4)\, \bar{\lambda}(x) \sigma_{\mu\nu} F^{\mu\nu} \lambda(x)$
\cite{Musberg:2013foa}.

\section{Baryon correlation functions}

\subsection{Baryon operators}

The mass of the lightest baryonic bound state in a channel specified by
particular quantum numbers is obtained from the correlation function
of a corresponding interpolating operator $W(x)$. Zero spatial momentum
is enforced by summing over spatial coordinates,
\begin{equation}
W_0(t) = \sum_{\vec{x}} W(t, \vec{x}).
\end{equation}
We consider local baryon operators $W(x)$ containing the product of three 
gluino fields $\lambda(x)$ at the same point $x$.
A possible general construction, similar to the Rarita-Schwinger field \cite{Rarita}, 
is
\begin{equation}
W(x) = t_{abc} \Gamma^A \lambda_a(x) \left( \lambda^T_b(x) \Gamma^B
\lambda_c(x) \right),
\end{equation}
where $\Gamma^A$ and $\Gamma^B$ are $4 \times 4$ spin matrices, and $W(x)$ is
a spinor. We choose $\Gamma^A=\mathbf{1}$ for simplicity, and denote
$\Gamma^B=\Gamma$.
In order that the baryon operator is a color singlet, $t_{abc}$ has to be an
invariant color tensor. One choice would be the completely antisymmetric
structure constants $f_{abc}$ of the gauge group. In the case of SU(2) this is
the antisymmetric tensor $\varepsilon_{abc}$. The matrix $\Gamma$ has then to be
symmetric, otherwise $W(x)$ would be zero identically due to the Grassmann nature
of the gluino field. For SU(3) there is another choice, namely the symmetric 
color tensor $d_{abc}$. In this case $\Gamma$ has to be antisymmetric.

The spin of the baryon depends on the choice of $\Gamma$.
Taking the Majorana condition
\begin{equation}
\bar{\lambda}(x) = \lambda^T(x) C
\end{equation}
into account, where $C$ is the charge conjugation matrix, the factor
$\lambda^T_b(x) \Gamma \lambda_c(x)$ transforms as a singlet under
spatial rotations for $\Gamma = C\gamma_4, C\gamma_5, \I \gamma_4C\gamma_5$.
Consequently, for these choices $W(x)$ describes a baryon with spin 1/2.
On the other hand, for $\Gamma = C\gamma_i$, $i=1,2,3$, the factor
$\lambda^T_b(x) \Gamma \lambda_c(x)$ transforms as a spatial vector, and
$W(x)$ will in general contain spin 3/2 and spin 1/2 
contributions \cite{GattringerLang}. The projections to definite spin are
involved and are discussed in \cite{Leinweber}.

\subsection{Baryonic correlation functions}

\begin{figure}[hbt!]
\centering
\includegraphics[width=0.5\textwidth]{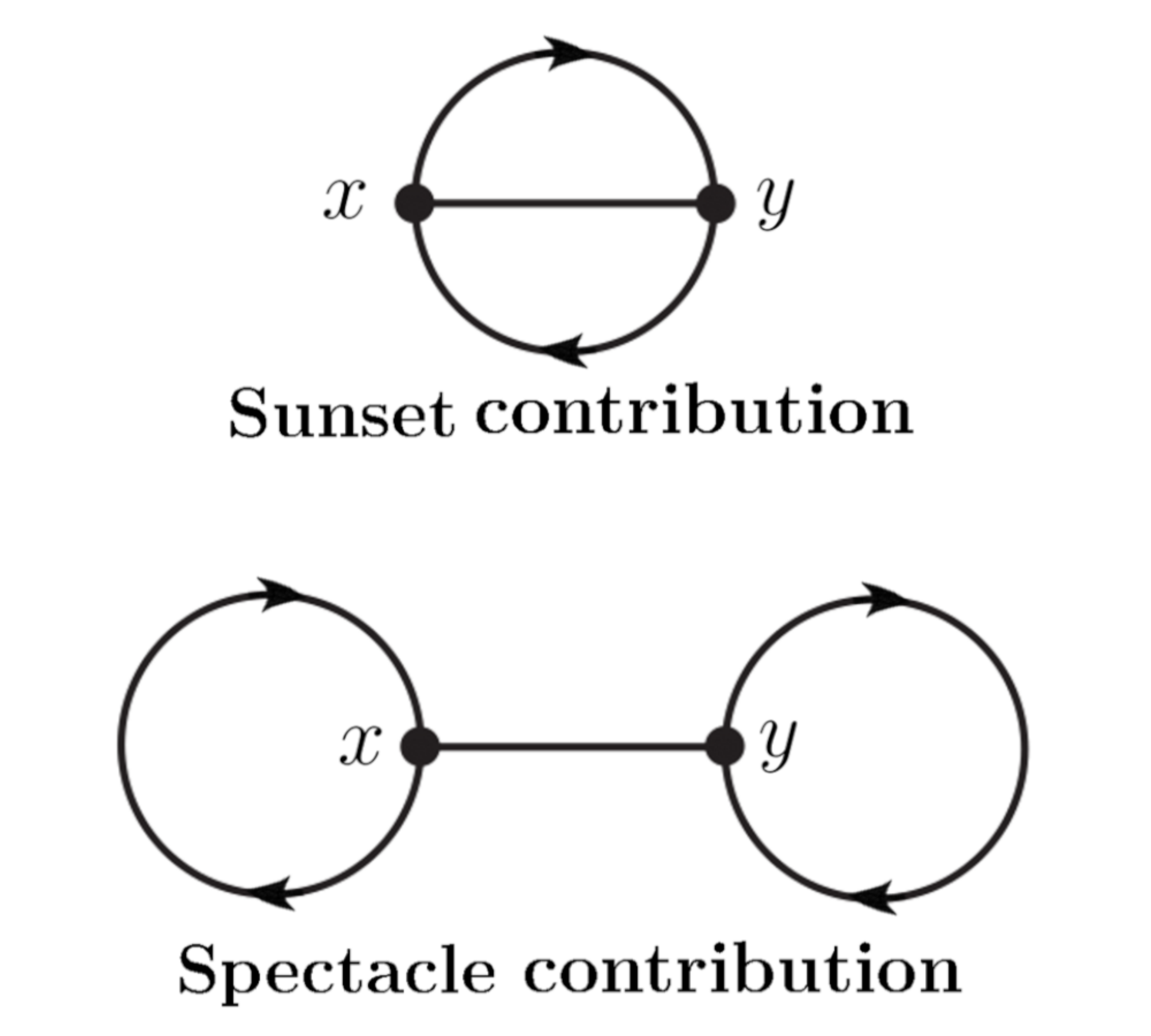}
\caption{The ``Sunset'' and ``Spectacle'' contributions to the baryon
correlation function in SYM theory.}
\label{SpectacleSunset}
\end{figure}

The correlation functions, needed for the computation of baryon masses,
are obtained from the interpolating field $W(x)$ and its conjugate field 
$\overline{W}(x)$ as
\begin{equation}
B(x,y) = \braket{W(x)\overline{W}(y)},
\end{equation}
where $\overline{W}(x)$ is given by
\begin{equation}
\overline{W}(x) = \left( C W(x) \right)^T
\end{equation}
up to a sign depending on the choice of the spin matrix \cite{Ali:2019thesis}.
With explicit Dirac indices the correlation function reads
\begin{align}
B^{\alpha \delta}(x,y)
& = \langle W^\alpha(x) \overline{W}^{\delta}(y) \rangle 
= \langle W^{\alpha}(x) C^{\delta \alpha'} W^{\alpha'}(y) \rangle \nonumber\\
& = t_{abc} t_{a'b'c'} \Gamma^{\beta\gamma} \Gamma^{\beta'\gamma'}
C^{\delta \alpha'}
\, \langle
\lambda_a^{\alpha}(x)\lambda_b^{\beta}(x)
\lambda_c^{\gamma}(x)\lambda_{a'}^{\alpha'}(y)\lambda_{b'}^{\beta'}(y)
\lambda_{c'}^{\gamma'}(y)\rangle.
\end{align}
In the numerical calculations the fermionic expectation values
\begin{equation}
\langle \mathcal{O} \rangle_F = \int \!\! D\lambda\ \mathcal{O}\, \E^{- S_f}
\end{equation}
in a given gauge field background are needed.
By Wick's theorem they can be expressed in terms of the gluino two-point function
\begin{equation}
K^{\alpha \beta}_{ab}(x,y) = 
\langle \lambda^{\alpha}_a(x) \lambda^{\beta}_b(y) \rangle_F .
\end{equation}
The antisymmetric matrix $K$ is related to the gluino propagator by
\begin{equation}
K^{\alpha\beta}_{ab}(x,y) = -\left(\Delta(x,y) C \right)^{\alpha\beta}_{ab},
\end{equation}
where the propagator $\Delta = D_w^{-1}$ is the inverse of the 
Wilson-Dirac matrix $D_w$.
For the product of six gluino fields, taking into account the fermionic signs,
we get the following 15 terms:
\begin{align}
\langle 
\lambda_a^{\alpha}(x) \lambda_b^{\beta}(x) \lambda_c^{\gamma}(x)
\lambda_{a'}^{\alpha'}(y) \lambda_{b'}^{\beta'}(y) &\lambda_{c'}^{\gamma'}(y) \rangle_F =\nonumber\\
& + K^{\alpha \beta}_{ab}(x,x)    K^{\gamma \alpha'}_{ca'}(x,y) K^{\beta' \gamma'}_{b'c'}(y,y) \nonumber\\
& - K^{\alpha \beta}_{ab}(x,x)    K^{\gamma \beta'}_{cb'}(x,y)  K^{\alpha'\gamma'}_{a'c'}(y,y) \nonumber\\
& + K^{\alpha \beta}_{ab}(x,x)    K^{\gamma \gamma'}_{cc'}(x,y) K^{\alpha'\beta'}_{a'b'}(y,y) \nonumber\\
& - K^{\alpha \gamma}_{ac}(x,x)   K^{\beta  \alpha'}_{ba'}(x,y) K^{\beta' \gamma'}_{b'c'}(y,y) \nonumber\\
& + K^{\alpha \gamma}_{ac}(x,x)   K^{\beta  \beta'}_{bb'}(x,y)  K^{\alpha'\gamma'}_{a'c'}(y,y) \nonumber\\
& - K^{\alpha \gamma}_{ac}(x,x)   K^{\beta  \gamma'}_{bc'}(x,y) K^{\alpha'\beta'}_{a'b'}(y,y) \nonumber\\
& + K^{\alpha \alpha'}_{aa'}(x,y) K^{\beta  \gamma}_{bc}(x,x)   K^{\beta' \gamma'}_{b'c'}(y,y) \nonumber\\
& - K^{\alpha \alpha'}_{aa'}(x,y) K^{\beta  \beta'}_{bb'}(x,y)  K^{\gamma \gamma'}_{cc'}(x,y) \nonumber\\
& + K^{\alpha \alpha'}_{aa'}(x,y) K^{\beta  \gamma'}_{bc'}(x,y)	K^{\gamma \beta'}_{cb'}(x,y) \nonumber\\
& - K^{\alpha \beta'}_{ab'}(x,y)  K^{\beta  \gamma}_{bc}(x,x)   K^{\alpha'\gamma'}_{a'c'}(y,y) \nonumber\\
& + K^{\alpha \beta'}_{ab'}(x,y)  K^{\beta  \alpha'}_{ba'}(x,y) K^{\gamma \gamma'}_{cc'}(x,y) \nonumber\\
& - K^{\alpha \beta'}_{ab'}(x,y)  K^{\beta  \gamma'}_{bc'}(x,y) K^{\gamma \alpha'}_{ca'}(y,y) \nonumber\\
& + K^{\alpha \gamma'}_{ac'}(x,y) K^{\beta  \gamma}_{bc}(x,x)   K^{\alpha'\beta'}_{a'b'}(y,y) \nonumber\\
& - K^{\alpha \gamma'}_{ac'}(x,y) K^{\beta  \alpha'}_{ba'}(x,y) K^{\gamma \beta'}_{cb'}(x,y) \nonumber\\
& + K^{\alpha \gamma'}_{ac'}(x,y) K^{\beta  \beta'}_{bb'}(x,y)  K^{\gamma \alpha'}_{ca'}(x,y) .
\end{align}
In the correlation function some of these terms can be combined, using the 
fact, that $t_{abc}$ is totally antisymmetric and $\Gamma^{\beta\gamma}$ is symmetric, 
or vice versa.
We are then left with
\begin{align}
B^{\alpha \delta}(x,y) = t_{abc} t_{a'b'c'} &\Gamma^{\beta\gamma} \Gamma^{\beta'\gamma'} C^{\delta\alpha'}
\nonumber\\
\{ & -2 K^{\alpha \alpha'}_{a a'}(x,y) K^{\beta  \beta'}_{bb'}(x,y)  K^{\gamma  \gamma'}_{cc'}(x,y)  \nonumber\\
& -4 K^{\alpha \beta'}_{a b'}(x,y)  K^{\beta  \gamma'}_{bc'}(x,y) K^{\gamma  \alpha'}_{ca'}(x,y)  \nonumber\\
& -2 K^{\alpha \beta}_{ab}(x,x)     K^{\gamma \alpha'}_{ca'}(x,y) K^{\gamma' \beta'}_{c'b'}(y,y)  \nonumber\\
& -4 K^{\alpha \beta}_{ab}(x,x)     K^{\beta' \gamma}_{b'c}(y,x)  K^{\gamma' \alpha'}_{c'a'}(y,y) \nonumber\\
& -1 K^{\alpha \alpha'}_{a a'}(x,y) K^{\beta  \gamma}_{bc}(x,x)   K^{\gamma' \beta'}_{c'b'}(y,y)  \nonumber\\
& +2 K^{\alpha \beta'}_{a b'}(x,y)  K^{\beta  \gamma}_{bc}(x,x)   K^{\gamma' \alpha'}_{c'a'}(y,y)
\}.
\end{align}
In the special case $\Gamma = C \gamma_4$, which we consider in our numerical work,
expressing the correlation function in terms of the propagator leads to
\begin{align}
B^{\alpha\alpha'}(x,y) = - f_{abc} f_{a'b'c'} &(C \gamma_4)^{\beta\gamma} (C \gamma_4)^{\beta'\gamma'}
\nonumber\\
\{ & + 2 \Delta^{\alpha \alpha'}_{a a'}(x,y)\Delta^{\beta \beta'}_{b b'}(x,y)\Delta^{\gamma \gamma'}_{c c'}(x,y)\nonumber\\
&  + 4 \Delta^{\alpha \beta'}_{a b'}(x,y) \Delta^{\beta \gamma'}_{b c'}(x,y) \Delta^{\gamma \alpha'}_{ca'}(x,y)\nonumber\\
&  + 2 \Delta^{\alpha \beta}_{ab}(x,x)   \Delta^{\delta \alpha'}_{ca'}(x,y)\Delta^{\delta' \beta'}_{c' b'}(y,y)C^{\gamma \delta}C^{\delta' \gamma'}\nonumber\\
&  + 4 \Delta^{\alpha \beta}_{ab}(x,x)   \Delta^{\beta' \gamma}_{b'c}(y,x) \Delta^{\gamma' \alpha'}_{c'a'}(y,y)\nonumber\\
&  + 1 \Delta^{\alpha \alpha'}_{aa'}(x,y)\Delta^{\beta \delta}_{bc}(x,x)   \Delta^{\delta' \beta'}_{c'b'}(y,y) C^{\gamma\delta} C^{\delta'\gamma'}\nonumber\\
&  + 2 \Delta^{\alpha \delta'}_{ac'}(x,y) \Delta^{\beta \delta}_{bc}(x,x)   \Delta^{\beta' \alpha'}_{b'a'}(y,y) C^{\gamma\delta} C^{\delta' \gamma'}
\}.
\label{baryonCorrFun}
\end{align}
With respect to the dependence on the space-time coordinates
the first two terms are summed up to the sunset contribution $B_{\textrm{Sset}}(x,y)$,
and the remaining four terms to the spectacle contribution $B_{\textrm{Spec}}(x,y)$, whose
graphical representations are given in Fig.~\eqref{SpectacleSunset}.

According to the availability of gauge ensembles and to obtain first
result for baryon masses, it is numerically less expensive
and convenient to consider gauge group SU(2). In this case the
baryon operator contains the antisymmetric structure constants
$t_{abc} = \varepsilon_{abc}$, and the spin matrix $\Gamma$ has to be
symmetric. We consider the choice $\Gamma = C\gamma_4$.\\

For zero momentum states, projection to definite parity can be accomplished with
the projection operators 
$P_{\pm} = \frac{1}{2} (1 \pm \gamma_4)$ \cite{MontvayMuenster}.
This finally gives
\begin{align}
B^{\pm}_{\textrm{Sset}}(x,y) = -\varepsilon_{abc} \varepsilon_{a'b'c'} &(C\gamma_4)^{\beta\gamma} (C\gamma_4)^{\beta'\gamma'} P_{\pm}^{\alpha\alpha'}
\nonumber\\
\langle &  + 2 \Delta^{\alpha \alpha'}_{a a'}(x,y)\Delta^{\beta \beta'}_{b b'}(x,y)\Delta^{\gamma \gamma'}_{c c'}(x,y)\nonumber\\
&  + 4 \Delta^{\alpha \beta'}_{a b'}(x,y) \Delta^{\beta \gamma'}_{b c'}(x,y) \Delta^{\gamma \alpha'}_{ca'}(x,y)
\rangle,
\end{align}
and
\begin{align}\label{Spec}
B^{\pm}_{\textrm{Spec}}(x,y) = -\varepsilon_{abc} \varepsilon_{a'b'c'} &(C\gamma_4)^{\beta\gamma} (C\gamma_4)^{\beta'\gamma'} P_{\pm}^{\alpha\alpha'} 
\nonumber\\
\langle &  + 2 \Delta^{\alpha \beta}_{ab}(x,x)   \Delta^{\delta \alpha'}_{c a'}(x,y)\Delta^{\delta' \beta'}_{c' b'}(y,y)C^{\gamma \delta}C^{\delta' \gamma'}\nonumber\\
&  + 4 \Delta^{\alpha \beta}_{a b}(x,x)   \Delta^{\beta' \gamma}_{b' c}(y,x) \Delta^{\gamma' \alpha'}_{c' a'}(y,y)\nonumber\\
&  + 1 \Delta^{\alpha \alpha'}_{a a'}(x,y)\Delta^{\beta \delta}_{b c}(x,x)   \Delta^{\delta' \beta'}_{c' b'}(y,y) C^{\gamma\delta} C^{\delta' \gamma'}\nonumber\\
&  + 2 \Delta^{\alpha \delta'}_{a c'}(x,y) \Delta^{\beta \delta}_{b c}(x,x)   \Delta^{\beta' \alpha'}_{b' a'}(y,y) C^{\gamma\delta} C^{\delta' \gamma'}
\rangle.
\end{align}
%

\section{Numerical results}

We have investigated the baryonic states in $\mathcal{N}=1$ supersymmetric Yang-Mills
theory with gauge group SU(2) by means of numerical Monte Carlo techniques. 
The correlation functions have been calculated based on configurations produced 
in previous work \cite{Bergner:2015adz,Bergner:2013nwa}.

As explained in the previous section, the baryon correlator consists of a sunset and a 
spectacle contribution that require different numerical methods. In both cases, the 
inverse of the Wilson-Dirac operator is required, which is provided by standard iterative 
solvers for a given input vector.

In the sunset contribution, all propagators connect the two lattice points and 
a point source can be chosen for the inversion. To complete the contractions, this
has to be repeated for all spin and color indices on the source side.

The spectacle part contains closed loop contributions ($\Delta(x,x)$ and $\Delta(y,y)$), 
in which the propagator connects each point with itself. These require techniques for a 
stochastic estimation of all-to-all propagators. We have already applied similar techniques  
for the estimation of mesonic operators in SYM. The inversion is done for several stochastic 
source vectors, which leads to an additional noise contribution in the signal. 
In practice we use 40 stochastic estimators combined with the exact contribution of 
the 200 lowest eigenmodes of $\gamma_5 D_W$.

The spectacle contribution combines the loops at $x$ and $y$ with a propagator. This is 
done by an inversion with a wall source vector at a time slice $x_0$ filled with appropriate 
entries from the stochastically estimated loop ($\Delta(x,x)$). The resulting sink vector is 
consequently contracted with the loop ($\Delta(y,y)$) at different time slices $y_0$.
The whole procedure is repeated for all source time slices $x_0$ to get the best signal for 
the average correlator $B(y_0,x_0)$.

\subsection{Discrete symmetries of the correlation functions}
\label{DisSymTest}

To cross-check the correctness of the numerical data for the correlation
function of Eq.~\eqref{baryonCorrFun}, discrete symmetries for time
reversal ($\mathcal{T}$) and parity ($\mathcal{P}$) are used
~\cite{Leinweber,Donini:1998}.
The baryon correlation function transforms according to
\begin{eqnarray}
\label{DP}
\!\!\!\!\!\!B(x,y)\!&\rightarrow&\!	B^{\mathcal{P}}(x^{\mathcal{P}},y^{\mathcal{P}}) \!=\! \braket{W^{\mathcal{P}}(x^{\mathcal{P}})\overline{W}^{\mathcal{P}}(y^{\mathcal{P}})} \!=\! \gamma_4B(x^{\mathcal{P}},y^{\mathcal{P}})\gamma_4,\\
\!\!\!\!\!\!B(x,y)\!&\rightarrow&\!	B^{\mathcal{T}}(x^{\mathcal{T}},y^{\mathcal{T}}) \!=\! \braket{W^{\mathcal{T}}(x^{\mathcal{T}})\overline{W}^{\mathcal{T}}(y^{\mathcal{T}})} \!=\! \gamma_4\gamma_5 B(x^{\mathcal{T}},y^{\mathcal{T}}) \gamma_5\gamma_4. \label{DT}
\end{eqnarray}
We consider the zero spatial momentum correlation function  
\begin{equation}
B(t) = \sum\limits_{\substack{\vec{x},\vec{y}
\\ t=x_0-y_0}} B(x,y).
\end{equation}
With the help of $\gamma_5$-hermiticity of the Wilson-Dirac matrix
we arrive at
\begin{align}
B_{\boldsymbol{1}}(t) &= - B_{\boldsymbol{1}}(N_t-t),\\
B_{\gamma_4}(t) &= B_{\gamma_4}(N_t-t),
\end{align}
where 
$B_{\boldsymbol{1}}(t) = \frac14 \tr\left[B(t)\right]$,
$B_{\gamma_4}(t) = \frac14 \tr\left[B(t)\gamma_4\right]$,
and $N_t$ is the time extent of the lattice.

\begin{figure}[hbt!]
\centering
\includegraphics[width=0.49\textwidth]{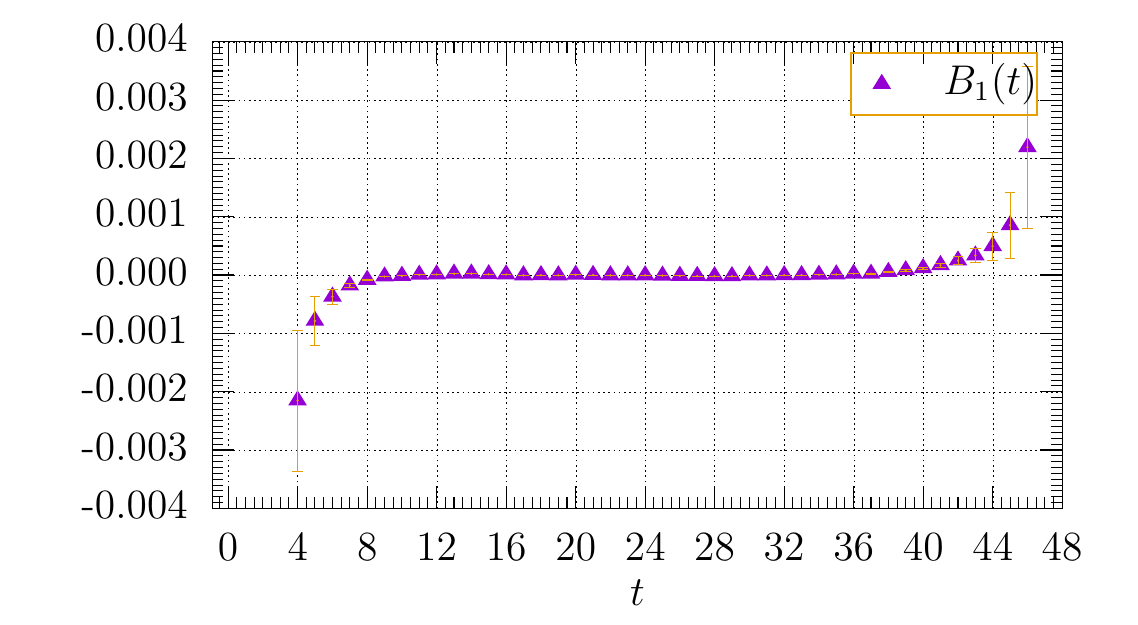}
\includegraphics[width=0.49\textwidth]{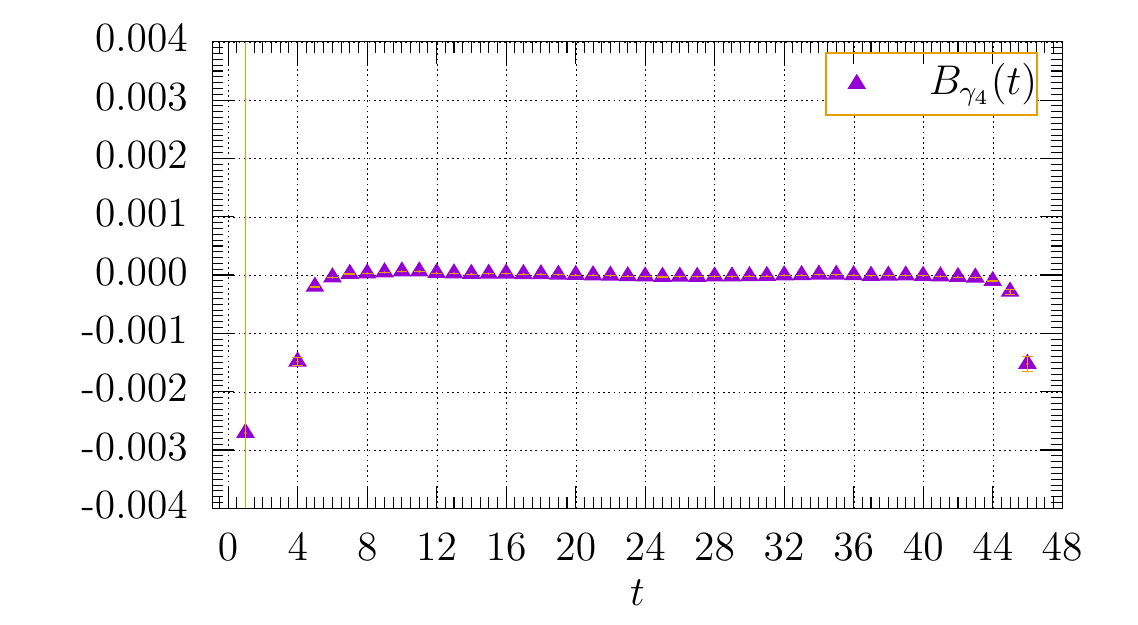}
\caption{Numerical results for the sunset contributions of the correlation
functions $B_{\boldsymbol{1}}(t)$ and $B_{\gamma_4}(t)$ 
at $\beta$=1.75 and $\kappa=0.14925$ for gauge group SU(2).}
\label{DSymTab}
\end{figure}

We have checked these symmetries for the sunset contributions, for which
much more precise numerical results are available compared to the spectacle 
contributions.
Fig.~\eqref{DSymTab} confirms that the sunset contribution of
$B_{\boldsymbol{1}}(t)$ is antisymmetric, and the one of
$B_{\gamma_4}(t)$ is symmetric within errors.
%

\subsection{Baryonic correlation functions and masses}

The numerical results of this exploratory study have been obtained 
for one ensemble of SU(2) SYM presented in \cite{Bergner:2015adz,Bergner:2013nwa}.
The lattice has size $24^3\times 48$, and the parameters are $\beta=1.75$
and $\kappa=0.14925$. A tree level Symanzik improved gauge action and a
Wilson-Dirac operator with one level stout smeared links has been applied.

\begin{figure}[hbt!]
	\centering
	\subfigure[Positive parity correlator.]{\label{Fig:corr_pp}
		\includegraphics[width=0.47\textwidth]{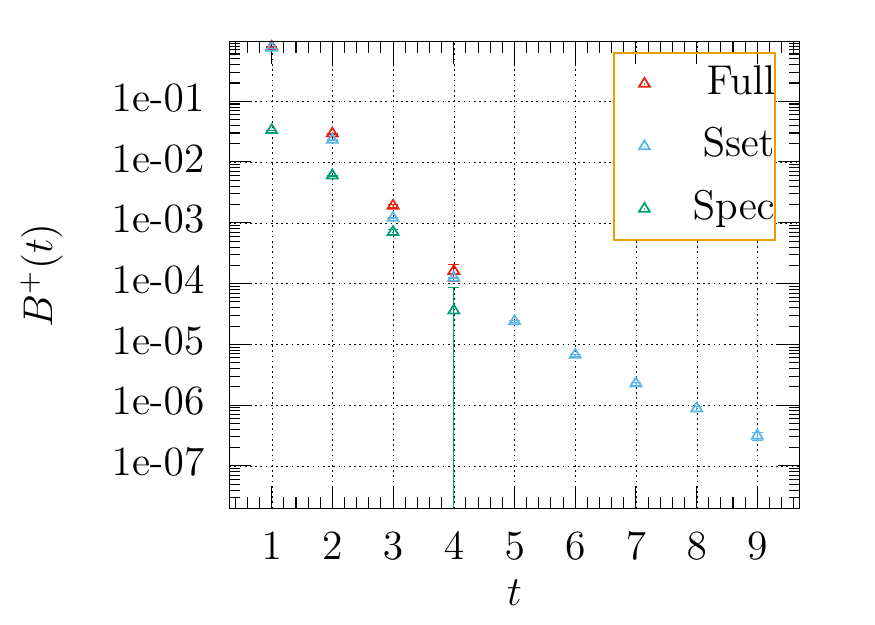}}
	\subfigure[Negative parity correlator.]{\label{Fig:corr_nn}
		\includegraphics[width=0.47\textwidth]{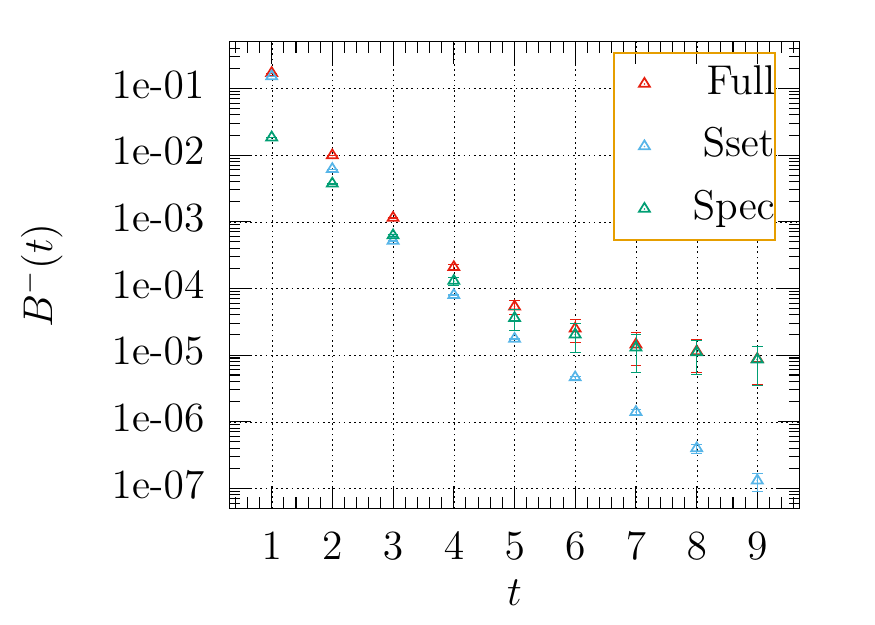}}
	\caption{Numerical results for the baryon correlation 
		functions for (a) positive parity and (b) negative parity at $\beta=1.75$ and 
		$\kappa=0.14925$ on a $24^3\times 48$ lattice. In addition to the full correlators, 
		the sunset (``Sset'') and spectacle (``Spec'') contributions are shown.}
	\label{PparNparCorr}
\end{figure}

The resulting propagators for positive and negative parity with their respective sunset 
and spectacle contributions are presented in Fig.~\ref{PparNparCorr}. 
A standard Jackknife procedure has been applied for error estimation.

The sunset contribution provides a much better signal than the spectacle one for all of 
the correlators. This contribution is similar to baryonic operators in QCD and hence an 
accuracy comparable with QCD data is achieved.
In SYM, however, the sunset contribution does not correspond to the correlator of a 
particle state. Only in a theory with a larger number of fermion species the sunset 
contribution is related to a physical bound state. In this sense the sunset contribution 
in SYM can be considered as a partially quenched approximation to a particle correlator.
The sunset contribution can be fitted quite accurately to a single exponential for both 
parities. The corresponding masses are rather large compared to the meson masses, 
see Tab.~\ref{tab:masses}.
\begin{table}[h!]
  \begin{center}
    \begin{tabular}{|c|c|c|c|c|c|}
    \hline
      \textbf{$am_{B^+_{\textrm{Sset}}}$} \rule[-2.0ex]{0pt}{3ex} & \textbf{$am_{B^-_{\textrm{Sset}}}$} & \textbf{$am_{B^-}$} & \textbf{$am_{\api}$} & \textbf{$a\mgg$} & \textbf{$am_{\aetap}$}\\
      \hline
              1.020(47)         &         1.207(82)         &     0.24(18)       &    0.20381(80)       &    0.3740(75)   &     0.299(28) \\
      \hline
    \end{tabular}
  \end{center}
    \caption{Masses of the baryon and two-particle bound states~\cite{Bergner:2015adz} 
    in $\mathcal{N}$=1 SUSY Yang-Mills theory with gauge group SU(2) for 
    $\beta=1.75$ and $\kappa=0.14925$.}
    \label{tab:masses}
\end{table}

The complete correlators are obtained by adding the spectacle contributions, which are much 
more noisy. The negative parity channel of the complete correlator provides a sufficient 
signal for an estimation the mass. An estimation of the positive parity mass has, however, 
not been possible with the current data.

The negative parity state appears to be significantly lighter than 
the one obtained considering only the sunset contribution.
The estimation of the lightest mass is in this case rather challenging since there seems to 
be a large excited state contribution, i.\,e.\ the prefactor of its exponential is 
quite large. 
In the following we explain the methods used to obtain the result for the negative parity 
state of Tab.~\ref{tab:masses}.
 
As a first test we have assumed a single exponential form of the correlator to obtain an 
effective mass at two lattice points. An estimate for the mass averaging these data in the 
range $t\in [6,8]$ would be around $m_{B^{-}}=0.31(35)$.
The masses obtained from a single exponential fit in different $t$-ranges are summarized in 
Tab.~\ref{tab:exp}. This mass estimate seems to decrease at larger $t$
towards values below $0.3$, but at the same time the signal gets overwhelmed by the error. 
A possible estimate is $m_{B^{-}}=0.31(18)$ from the fit interval $t\in [6,9]$.  This is an 
indication that excited state contamination is rather large at the accessible $t$-range of 
the correlator.
\begin{table}[h!]
  \begin{center}
    \begin{tabular}{|c|c|c|c|c|}
    \hline
      Fit range ($t$) & $A$ & $\sigma_{A}$ & $m$ & $\sigma_{m}$ \\
      \hline
6--8 & 1.54856e-04  & 2.76203e-04  & 0.32236   & 0.25463  \\
6--9 & 1.42584e-04  & 2.00976e-04  & 0.31265   & 0.18452  \\
7--9 & 9.98849e-05  & 2.17467e-04  & 0.26426   & 0.28608  \\
      \hline
    \end{tabular}
  \end{center}
    \caption{Estimates of the mass $m$ and multiplicative factor $A$ for negative parity
    using the fit function $A\,\E^{-mt}$. The errors are denoted by $\sigma$.}
    \label{tab:exp}
\end{table}

In order to remove the excited state contamination, we have done double exponential fits. The results in Tab.~\ref{tab:exp2m} and \ref{tab:exp2A} show that this provides more consistent data even at smaller $t$ ranges compared to the single exponential fit. A mass estimate is $m_{B^{-}}=0.24(18)$ from the fit interval $t\in [3,9]$.
As can be seen from the values of the prefactor $A_1$ in Tab.~\ref{tab:exp2A}, there is a large contribution from the heavy mass of the excited state.
\begin{table}[h!]
  \begin{center}
    \begin{tabular}{|c|c|c|c|c|}
    \hline
      Fit range ($t$) & $m_1$ & $\sigma_{m_1}$ & $m$ & $\sigma_{m}$ \\
      \hline
3--7 & 1.82570   & 0.12008   & 0.28226   & 0.42181  \\
3--8 & 1.80280   & 0.09764   & 0.21303   & 0.24616  \\
3--9 & 1.81288   & 0.09521   & 0.24190   & 0.17894  \\
4--8 & 1.69159   & 0.65843   & 0.13819   & 0.58371  \\
4--9 & 1.78465   & 0.57213   & 0.22977   & 0.29006  \\
      \hline
    \end{tabular}
  \end{center}
    \caption{Masses are estimated by fitting the function 
    $A_1\,\E^{-m_1t} + A\,\E^{-mt}$ to the correlator data for different fit 
    ranges (negative parity).
    \label{tab:exp2m}}
\end{table}
\begin{table}[h!]
  \begin{center}
    \begin{tabular}{|c|c|c|c|c|}
      \hline
      Fit range ($t$) & $A_1$ & $\sigma_{A_1}$ & $A$ & $\sigma_{A}$ \\
      \hline
3--7 & 2.61276e-01  & 8.32151e-02  & 1.10252e-04  & 3.40647e-04  \\
3--8 & 2.46011e-01  & 6.60862e-02  & 6.56622e-05  & 1.16709e-04  \\
3--9 & 2.52735e-01  & 6.70216e-02  & 7.83048e-05  & 1.09974e-04  \\
4--8 & 1.79004e-01  & 5.58144e-01  & 4.40431e-05  & 1.71411e-04  \\
4--9 & 2.43887e-01  & 6.00709e-01  & 7.41955e-05  & 1.71981e-04  \\
      \hline
    \end{tabular}
  \end{center}
    \caption{The parameters $A$ and $A_1$ for the fits in Tab.~\ref{tab:exp2m}.
    \label{tab:exp2A}}
\end{table}

We have tested further methods like a fit of the excited state contamination using only the 
sunset part, but without reasonable improvement.
Our final best estimates in Tab.~\ref{tab:masses} have been obtained from a multistate fit 
analysis which uses the $\cosh$ function and Akaike information criterion (AIC) explained 
in~\cite{Bazavov:2019msm} and references therein.
Note that all methods provide results consistent within the errors.

\section{Conclusion}
We have presented a discussion of baryonic bound states in $\mathcal{N}=1$ supersymmetric 
Yang-Mills theory. It is usually not expected that these are part of the lightest multiplets 
of the theory. These states are similar to baryonic states of QCD, but their correlators have 
a different type of contractions and require a spectacle contribution in addition to the
usual sunset diagrams.

We have done a first exploratory numerical study of correlators and particle masses for these 
bound states. The sunset contribution alone leads to a rather heavy particle mass. It is 
quite challenging to provide a reasonable result including the spectacle contribution due to 
the small signal to noise ratio. Our first estimates suggest a mass in the negative parity 
channel which is compatible with the lightest multiplet. This might be due to an overlap with 
the gluino-glue bound state, which is the fermionic member of the lightest multiplet. 

Further improvements of the measurement are possible. The most relevant one is a detailed 
analysis of smearing methods to reduce the overlap with excited states. We plan to test this 
in a subsequent analysis of the SYM spectrum.

\section*{Acknowledgments}

We thank Henning Gerber and Philipp Scior for many helpful discussions
and aid with the numerical work.
The authors gratefully acknowledge the Gauss Centre for Supercomputing
e.\,V.\,(www.gauss-centre.eu) for funding this project by providing
computing time on the GCS Supercomputer JUQUEEN and JURECA at J\"ulich
Supercomputing Centre (JSC) and SuperMUC at Leibniz Supercomputing Centre
(LRZ). Further computing time has been provided on the compute cluster PALMA
of the University of M\"unster. This work is supported by the Deutsche
Forschungsgemeinschaft (DFG) through the Research Training Group ``GRK 2149:
Strong and Weak Interactions - from Hadrons to Dark Matter''.
G.~B.\ is funded by the Deutsche Forschungsgemeinschaft (DFG) under Grant 
Nos.~432299911 and 431842497.
S.~Ali acknowledges financial support from the Deutsche
Akademische Austauschdienst (DAAD).


\end{document}